\documentclass[showpacs,aps,prl,twocolumn,nofootinbib]{revtex4}
\usepackage{amsmath}
\usepackage{epsfig}
\usepackage{subfigure}
\usepackage{float}
\usepackage{verbatim}

\newcommand{\ie}{{i.e. }}

\newcommand{\cf}[1]{{Fig.~\ref{#1}}}

\begin{document}
\leftline{} \rightline{CP3-08-17,DSF-16/08}
\title{$\Upsilon$ production at the Tevatron and the LHC}

\author{P.~Artoisenet$^{a}$, J.~Campbell$^{b}$, J.P.~Lansberg$^{c}$, F.~Maltoni$^{a}$, F.~Tramontano$^{d}$}
\affiliation{
$^{a}$Center for Particle Physics and Phenomenology (CP3), Universit\'e catholique de Louvain, B-1348 
Louvain-la-Neuve, Belgium\\
$^{b}$Department of Physics and Astronomy, University of Glasgow, 
Glasgow G12 8QQ, United Kingdom \\
$^{c}$ Institut f\"ur Theoretische Physik, Universit\"at Heidelberg, D-69120 
Heidelberg, Germany\\
$^{d}$ Universit\`a di Napoli Federico II, Dipartimento di Scienze Fisiche, and INFN, Sezione di Napoli, I-80126 Napoli, Italy
}


\begin{abstract}
We update the theoretical predictions for direct $\Upsilon(nS)$
hadroproduction in the framework of NRQCD. We show that the
next-to-leading order corrections in $\alpha_S$ to the color-singlet
transition significantly raise the differential cross section at high
$p_T$ and substantially affect the polarization of the
$\Upsilon$. Motivated by the remaining gap between the NLO yield and
the cross section measurements at the Tevatron, we evaluate the
leading part of the $\alpha^5_S$ contributions, namely those coming
from $\Upsilon(nS)$ associated with three light partons. The
differential color-singlet cross section at $\alpha^5_S$ is in
substantial agreement with the data, so that there is no evidence for
the need of color-octet contributions. Furthermore, we find that the
polarization of the $\Upsilon(nS)$ is longitudinal. We also
present our predictions for $\Upsilon(nS)$ production at the LHC.
\end{abstract}


\maketitle


Even though the discrepancy between experimental measurements and
theoretical predictions for $\Upsilon(nS)$ hadroproduction is less
dramatic than for $J/\psi$ and $\psi'$, it is still unclear which are
the dominant mechanisms at work in those processes~\cite{review}.  The
leading order (LO) $\alpha_S^3$ result~\cite{CSM_hadron} 
in the $v$ expansion of non-relativistic QCD (NRQCD)~\cite{Bodwin:1994jh} neither predicts the yield seen in the data~\cite{Affolder:1999wm,Acosta:2001gv},
undershooting it by a factor of ten at $p_T$ around 20 GeV, nor shows
the correct $p_T$ behavior.  In this case it has been shown that the
inclusion of higher-$v$ corrections, in particular those associated
with the color-octet mechanism, allows the data to be described.
The yield and the $p_T$ shape can be reproduced, at the cost 
of introducing two unknown long-distance-matrix elements of
NRQCD that can be determined from fits to the data~\cite{Braaten:2000cm}. 
However, the polarization predicted following this approach disagrees 
with both the published CDF~\cite{Acosta:2001gv} and new preliminary 
D$\emptyset$ measurements~\cite{D0note}.

In view of such a puzzling situation, it is worth reexamining in
detail the available theoretical predictions.  In
Refs.~\cite{Campbell:2007ws,Artoisenet:2007xi}, the complete set of
$\alpha_S^4$ corrections to color-singlet production was calculated
for the first time. The results confirmed the importance of the new
channels that open up at this order, which is due to their different
$p_T$ scaling (see \cf{diagrams} (c,d)).  In this Letter we
apply those results to a phenomenological study of $\Upsilon$
production at the Tevatron, comparing the differential cross section
at next-to-leading order (NLO) with the data from the CDF
collaboration.  We also argue that some $\alpha^5_S$ contributions
coming from three-jet configurations, \ie $\Upsilon + jjj$, such as
those arising from gluon-fragmentation (\cf{diagrams} (e)) and
``high-energy enhanced'' (\cf{diagrams} (f)) channels are
important. We show that by adding such $\alpha_S^5$ contributions to
the NLO yield, the experimental measurements are reproduced.  We also
predict the polarization of the $\Upsilon$ from the angular
distribution of the leptons produced in the decay.

\begin{figure}[ht!]
\centering
\subfigure[]{\includegraphics[scale=.33]{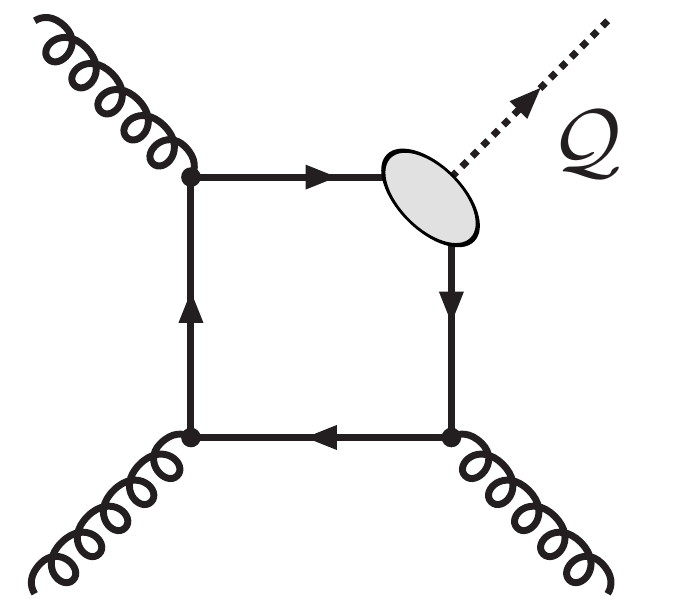}}
\subfigure[]{\includegraphics[scale=.33]{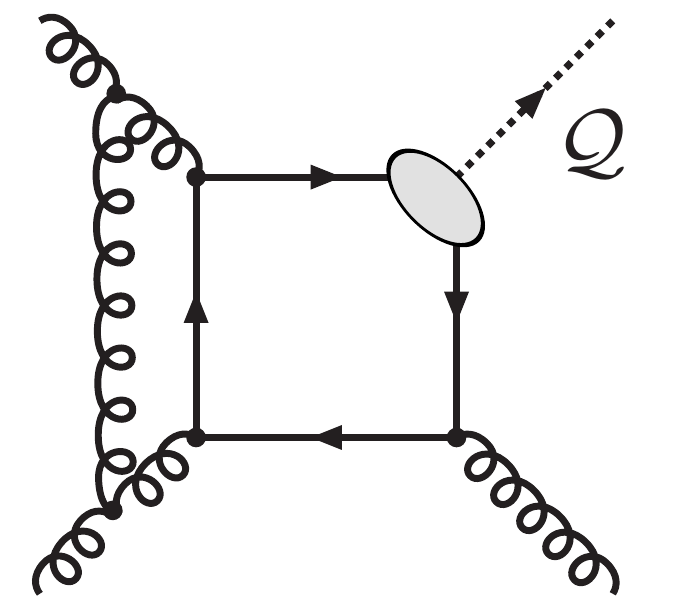}}
\subfigure[]{\includegraphics[scale=.33]{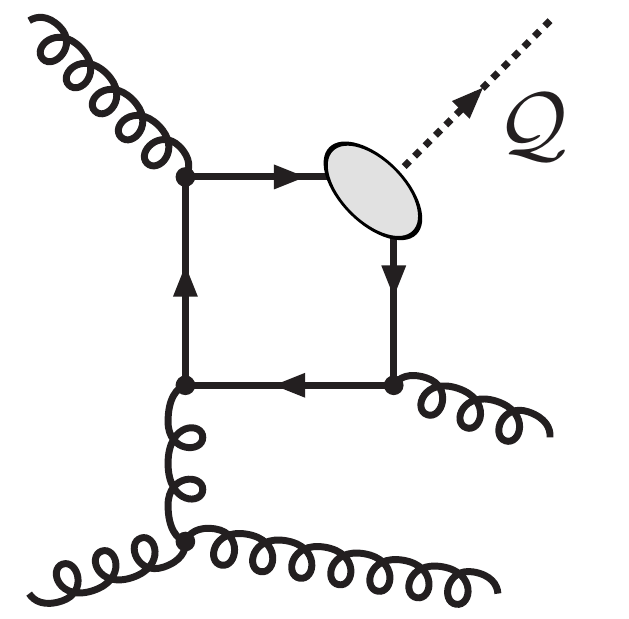}}
\subfigure[]{\includegraphics[scale=.33]{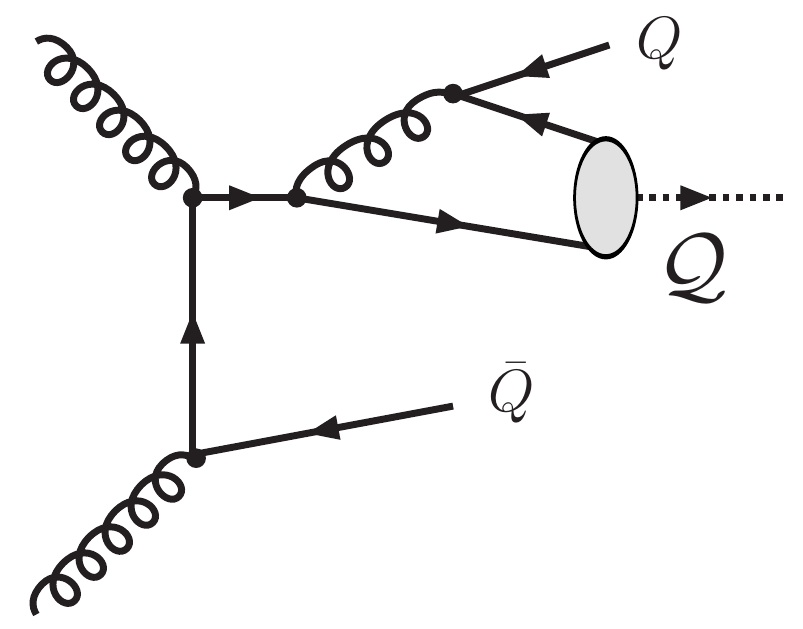}}
\subfigure[]{\includegraphics[scale=.33]{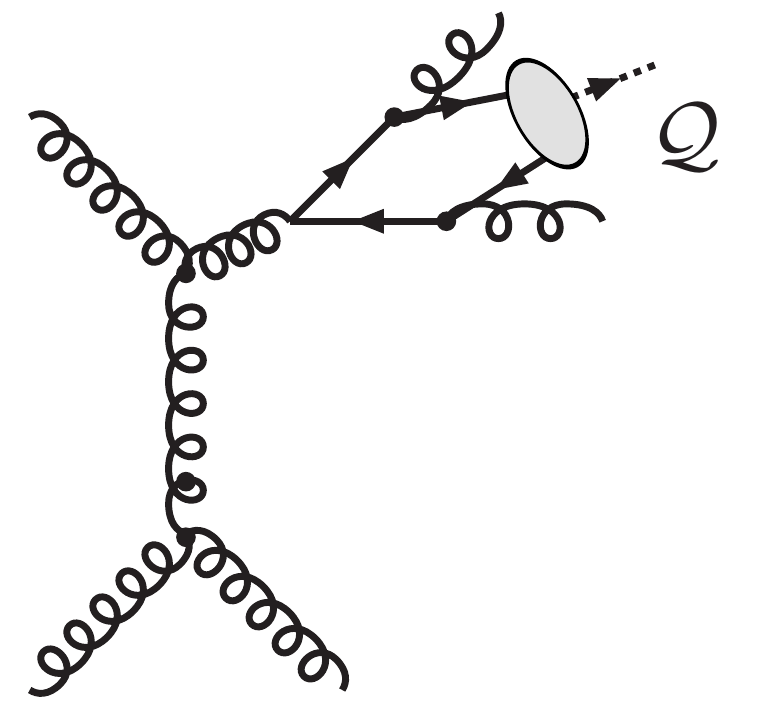}}
\subfigure[]{\includegraphics[scale=.33]{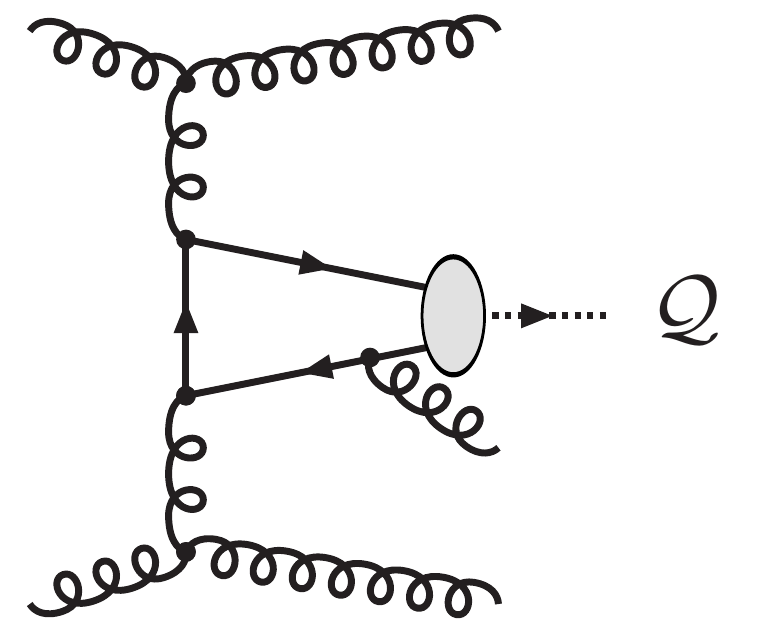}}
\caption{Representive diagrams contributing to $\Upsilon$ hadroproduction
at orders $\alpha_S^3$ (a), $\alpha_S^4$ (b,c,d), $\alpha_S^5$ (e,f). See discussions in the text.}
\label{diagrams}
\end{figure}


We first present the NLO predictions for the $p_T$ spectrum of the
$\Upsilon$ at leading order in $v$, following the same procedure
as in Ref.~\cite{Campbell:2007ws} and
adding information on the polarization of the quarkonium state.  
In our numerical evaluation of the differential cross section and polarization parameters, we have
used: $|R_{\Upsilon(1S)}(0)|^2=6.48$ GeV$^3$ and
$|R_{\Upsilon(3S)}(0)|^2=2.47$ GeV$^3$~\cite{review}; $\mu_0=\sqrt{(2
m_b)^2+p_T^2}$; Br$(\Upsilon(1S) \to \mu^+ \mu^- ) = 0.0248$ and
Br$(\Upsilon(3S) \to \mu^+ \mu^- ) = 0.0218$; $m_b=4.75$ GeV; PDF set:
CTEQ6M~\cite{Pumplin:2002vw}. For comparison, we also plot the LO
yield, for which we used the PDF set CTEQ6L1.

The experimental results for the cross section differential in $p_T$
for prompt $\Upsilon(1S)$ and $\Upsilon(2S)$, as well as the cross section for direct
$\Upsilon(3S)$ and the polarization for prompt $\Upsilon(1S)$ were obtained in Run I at
$\sqrt{s}=$1.8 TeV~\cite{Acosta:2001gv}. For Run II at $\sqrt{s}=$1.96 TeV, so far
only D$\emptyset$ have analyzed data for $\Upsilon(1S)$~\cite{Abazov:2005yc}.
The comparison performed in~\cite{Abazov:2005yc} shows full agreement
with~\cite{Acosta:2001gv}, once  the difference in
$\sqrt{s}$ is taken into account.

In order to compare our calculation for direct $\Upsilon(1S)$ production
with the CDF data, we have multiplied the most recent
prompt-$\Upsilon(1S)$ cross-section measurement
of~\cite{Acosta:2001gv} by the direct fraction
$F^{\Upsilon(1S)}_{\textrm{direct}}=0.5 \pm 0.12$. This fraction was
obtained from an older sample~\cite{Affolder:1999wm}, where the
selection cuts $|p_T|>8$ GeV and $|\eta|<0.7$ were applied. The
similar $p_T$ dependence observed for prompt-$\Upsilon(1S)$ and
direct-$\Upsilon(3S)$ production, as shown in Fig. 2
of~\cite{Acosta:2001gv}, confirms that this fraction should not depend
very strongly on $p_T$. This justifies using the same fraction
for the whole $p_T$ range.  The errors from the prompt cross section
and the direct fraction have been  combined in quadrature. 
Similar information on the direct yield of  $\Upsilon(2S)$ is not yet available.

Our results for direct production of $\Upsilon(1S)$ and $\Upsilon(3S)$
at $\sqrt{s}=1.8$ TeV are shown in~\cf{upsilon1S1800}. The curves
labeled as NNLO$^\star$ will be discussed later on.  In
order to estimate the theoretical uncertainties, we have varied the
renormalization and factorization scales between $0.5 \mu_0$ and $2
\mu_0$, keeping them equal. We have combined the resulting
uncertainties in quadrature with those resulting from variation of the bottom
quark mass, $m_b=4.75 \pm 0.25$ GeV.  The uncertainty associated with the non-perturbative parameter $|R_{\Upsilon(nS)}(0)|^2$~\cite{Braaten:2000cm,Bodwin:2007fz}, corresponding to an overall normalization, is not included.  The curves for direct $\Upsilon(2S)$ production can be easily obtained -- at this order in
$v$ -- by adopting the corresponding values for the leptonic branching
ratio and for the non-perturbative matrix element,
$|R_{\Upsilon(2S)}(0)|^2$.

\begin{figure}[ht!]
\centering
{\includegraphics[width=0.97\columnwidth,angle=0]{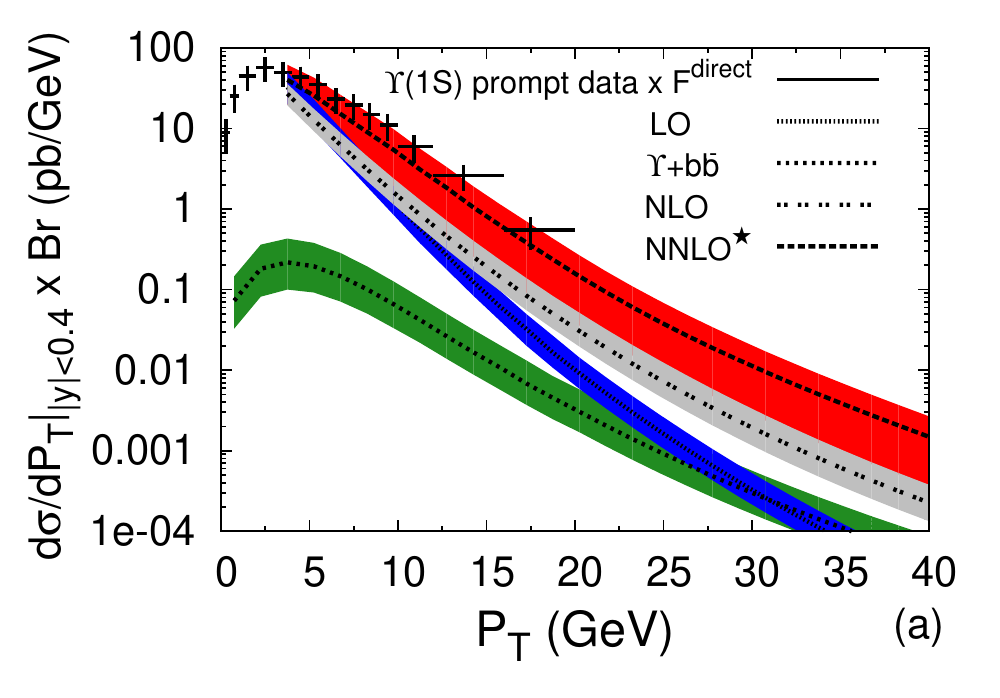}}
{\includegraphics[width=0.97\columnwidth,angle=0]{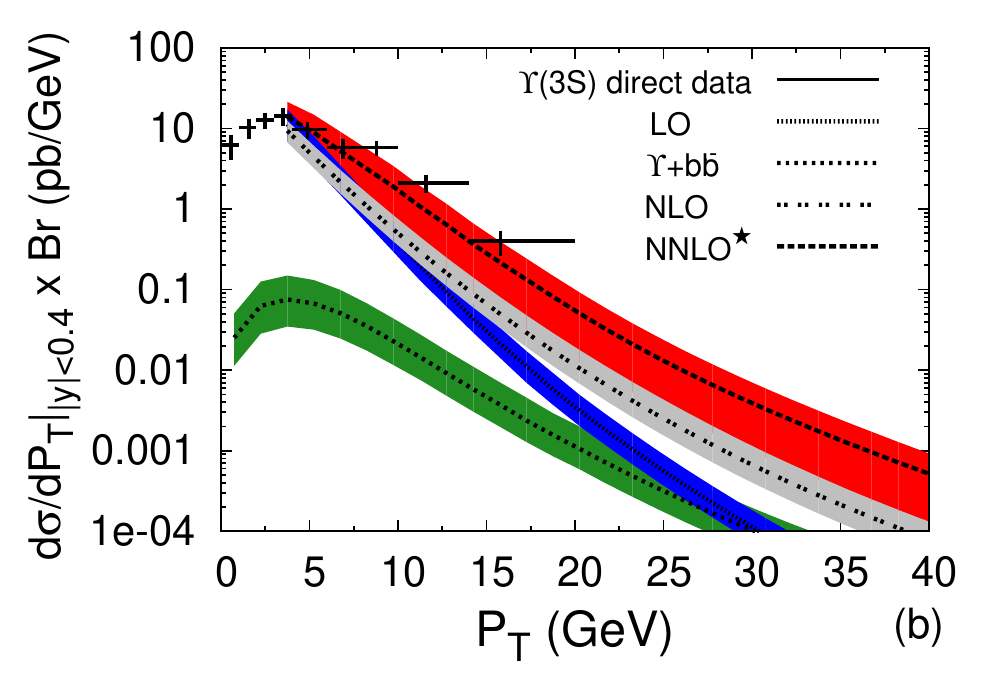}}
\caption{Results for (a) $\Upsilon(1S)+X$ and (b) $\Upsilon(3S)+X$ at LO ($\alpha_S^3$), 
for associated production ($\alpha_S^4$), 
for full NLO ($\alpha_S^3$+$\alpha_S^4$) and for NNLO$^\star$ (up to $\alpha_S^5$) compared with the 
 direct yield at $\sqrt{s}=1.8$ TeV measured by the CDF collaboration~\cite{Acosta:2001gv}. 
The theoretical-error bands 
for LO, NLO and associated production come from combining the
uncertainties resulting from the choice of $\mu_f$, $\mu_r$, $m_b$. The uncertainty of NNLO$^\star$ includes
the variation of the cutoff, $s_{ij}^{\rm min}$,  between  $0.5 m_b^2$ and $2 m_b^2$  and of the scales $\mu_r$ and $\mu_f$.}
\label{upsilon1S1800}
\end{figure}

\begin{figure}[ht!]
\centering
\includegraphics[width=\columnwidth]{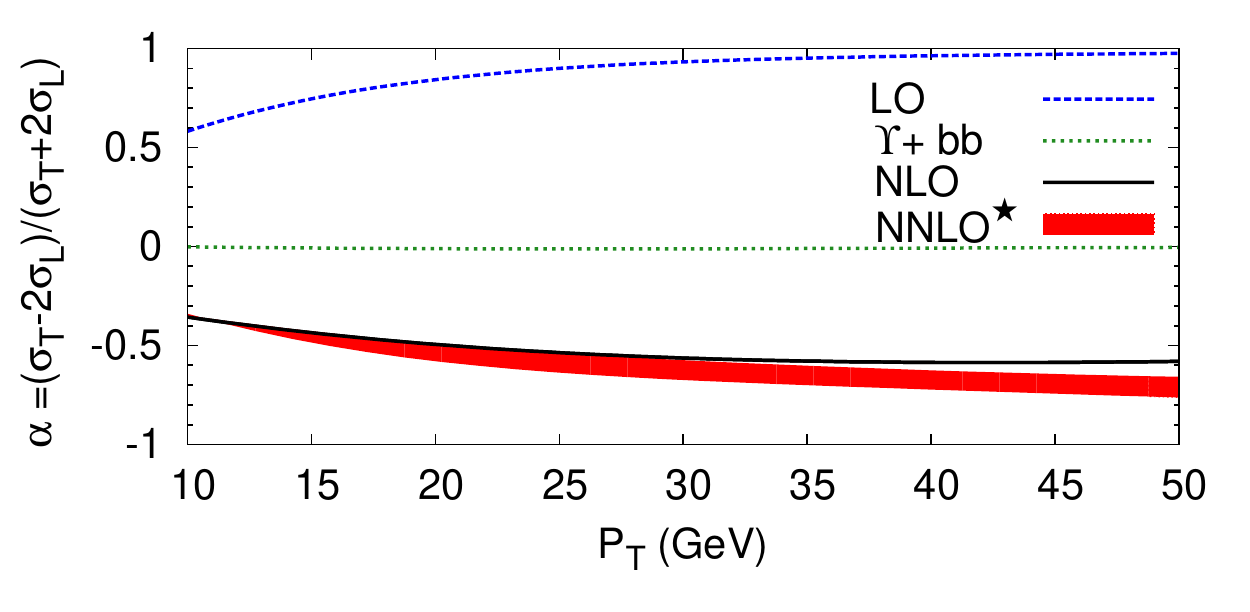}
\caption{Polarization for direct 
$\Upsilon(nS)$ ($n=1,2,3$) production at $\sqrt{s}=1.8$ TeV up to the order $\alpha_S^5$ (NNLO$^\star$). Most of the 
uncertainties on $\alpha$ for LO, associated production and NLO cancel. The uncertainty band of the (NNLO$^\star$) result comes from the variation of the
cutoff $s_{ij}^{\rm min}$. }
\label{polarizationalphas4}
\end{figure}

As a first comment, we note that the LO and NLO results are very
similar at low $p_T$, \ie where the bulk of the cross section is
found. This implies that the $\alpha_S^4$ corrections are small
in this region and that the perturbative expansion of the total cross
section is under control. On the other hand, the relative importance
of the  NLO corrections dramatically increases at larger $p_T$, where new
$p_T^{-6}$ channels (\cf{diagrams} (c)) start to dominate. This
behavior significantly reduces the discrepancy with the
experimental data that is found at LO.  However, the NLO prediction still drops
too fast in comparison with the data.  Note that the NLO calculation also takes
into account associated production $\Upsilon +b\bar b$ (\cf{diagrams} (d)).
This contribution, in spite of reproducing the shape
of the data very well at both low and high $p_T$ (asymptotically
it behaves as $p_T^{-4}$ ), remains negligible up to at least $p_T=30$ GeV.

Throughout our calculation, the polarization information of the
$\Upsilon$ can be traced through the angular distributions of the two decay
leptons.  Defining $\theta$ as the angle between the $\ell^+$ direction in the
quarkonium rest frame and the quarkonium direction in the laboratory
frame, the normalized angular distribution $I(\cos \theta)$ reads
\begin{equation}
\label{angulardist}
I(\cos \theta) =
\frac{3}{2(\alpha+3)} (1+\alpha \, \cos^2 \theta)\,,
\end{equation}
from which we have extracted the polarization parameter $\alpha$ bin by bin in $p_T$.
Our results are shown in~\cf{polarizationalphas4}
along with the curves for the LO yield and for the associated production alone. 
Even though the inclusion of NLO corrections brings the predictions 
closer to the prompt data, the discrepancy in the 
normalization makes such a comparison not very compelling.


As already noted, the discrepancy between the data and the full NLO
(up to $\alpha^4_S$) result in~\cf{upsilon1S1800} increases with
$p_T$. This might be taken as an indication that one or more processes
with a more gentle $p_T$ scaling than the $\alpha^4_S$ contributions,
which scale at most as $p_T^{-6}$ (disregarding the sub-dominant
$\Upsilon + b \bar b$ channel), are still missing.  In this section,
we argue that the $\alpha_S^5$ terms provide us with the missing
contributions.  Indeed, the gluon fragmentation channel (\cf{diagrams}
(e)) occurs for the first time at order $\alpha_S^5$ and behaves like
$p_T^{-4}$ at large transverse momentum. At next-to-next-to-leading
order (NNLO) ``high-energy enhanced'' channels (\cf{diagrams} (f))
also appear for the first time, whose contributions have previously
been approximated in the $k_T$ factorization approach~\cite{kt}. Both
these contributions, which exhaust the possibilities of new
kinematical enhancements at higher order, are at the Born level and
therefore finite at $\alpha_S^5$. They are a subset of $p\bar p\to
\Upsilon + jjj$ ($j$ stands for $u,d,s,c,g$) and provide us with new
mechanisms to produce a high-$p_T$ $\Upsilon$ with a lower kinematic
suppression. They can therefore be expected to dominate the
differential cross section at NNLO accuracy in the region of large
transverse momentum.

We present here an estimate of these dominant $\alpha_S^5$
contributions at large $p_T$, bypassing the presently out-of-reach
inclusive calculation of $p\bar p\to \Upsilon + X$ at NNLO accuracy.
We consider the whole set of $\alpha_S^5$ processes contributing to
the color-singlet production of a $\Upsilon$ associated with three light
partons. The computation of the associated matrix elements is
technically possible via the approach presented in
Ref.~\cite{Artoisenet:2007qm}.  In order to protect the integrated
cross section from soft and collinear divergences, we impose a
``democratic'' invariant-mass cut $s_{ij}^{\rm min}$ of the order of $m_b^2$ to
any pair of light partons so that the phase-space integration remains
finite.  For new channels opening up at $\alpha_S^5$, the dependence
on the value of $s_{ij}^{\rm min}$ is expected to be small in the
region $p_T\gg m_b$, since no collinear or soft divergences can appear
there.  For channels whose Born-level contribution occurs at
$\alpha_S^3$ or $\alpha_S^4$, the invariant-mass cut results in
potentially large logarithms of $s_{ij}/s_{ij}^{\rm min}$ 
in the differential cross sections. However, these logs
factorize over lower order amplitudes that are suppressed by powers of
$p_T$ compared to the dominant $\alpha_S^5$ contributions.  The result
is that the sensitivity to the value chosen for the minimal invariant
mass cut $s_{ij}^{\rm min}$ is expected to die away as $p_T$ increases.

In fact, this argument can be tested by using it to estimate the
dominant contribution to the production of $\Upsilon$ at NLO
accuracy. The differential cross section for the real $\alpha_s^4$
corrections, $\Upsilon+jj$ production, is displayed in
\cf{alphas4real}. The grey band is obtained by varying the
invariant-mass cut $s_{ij}^{\rm min}$ between any pairs of light partons from
$0.5 m_b^2$ to $2m_b^2$.  The yield becomes insensitive to the value
of $s_{ij}^{\rm min}$ as $p_T$ increases, and it reproduces very accurately the
differential cross section at NLO accuracy.
 
\begin{figure}[ht!]
\centering \includegraphics[width=0.95\columnwidth]{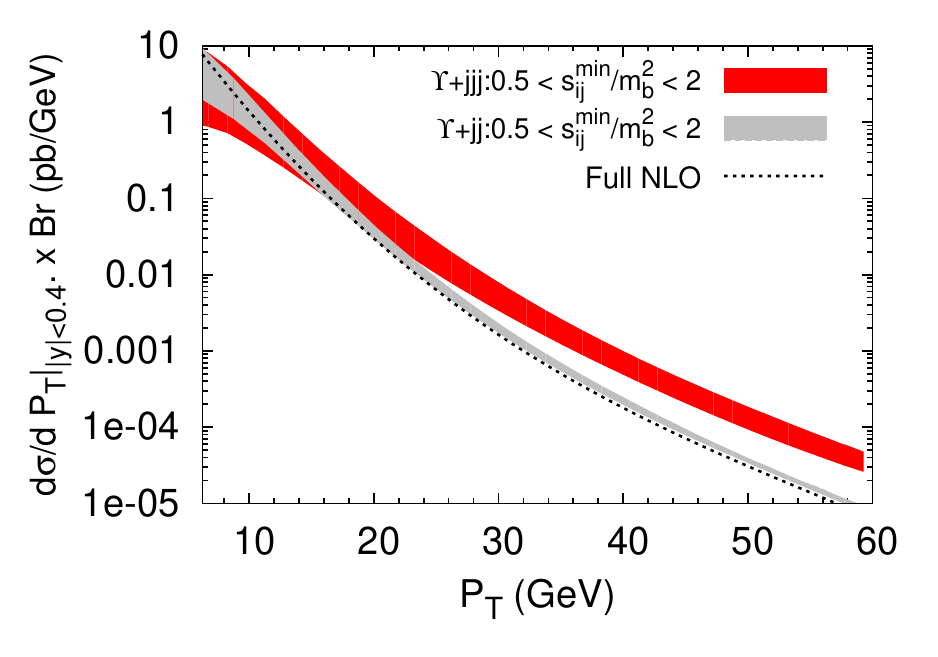}
\caption{Full computation at NLO for $\Upsilon + X$ (dashed line)  vs. 
$\Upsilon(1S)$ + 2 light partons with a cut on $s_{ij}^{\rm min}$ (grey band). 
The red band shows the cross section for $\Upsilon(1S)$ + 3 light 
partons with the same cut. (see text for details).
}
\label{alphas4real}
\end{figure}

The computation of $p \bar p \rightarrow \Upsilon +jjj$ at tree level 
is in principle straightforward, but technically quite challenging: 
several parton-level subprocesses contribute, each
involving up to hundreds of Feynman diagrams.  We follow 
the approach described in Ref.~\cite{Artoisenet:2007qm}, which
allows the automatic generation of both the subprocesses and 
the corresponding scattering amplitudes. We find that the subprocess 
$gg \to \Upsilon+ggg$ dominates, providing  50\% of the whole
yield.  The uncertainty associated with the choice of the cut 
$s_{ij}^{\rm min}$ is substantial although, as can be seen in 
\cf{alphas4real}, it decreases at large $p_T$ where these 
contributions are dominant.

The differential cross-sections for $\Upsilon(1S)$ and $\Upsilon(3S)$
are shown in~\cf{upsilon1S1800}.  The red band (referred to as
NNLO$^\star$) corresponds to the sum of the NLO yield and the
$\Upsilon+jjj$ contributions.  The contribution of $\Upsilon$
production with three light partons fills the gap between the data and
the NLO calculation. The $\alpha_S^5$ contribution is very sensitive
to the choice of the renormalization scale, $\mu_r$.  This is
expected: for moderate values of the $p_T$, the missing virtual part
might be important, whereas at large $p_T$, the yield is dominated by
Born-level $\alpha_S^5$-channels from which we expect a large
dependence on $\mu_r$. Even though the uncertainty on the
normalization is rather large, the prediction of the $p_T$ shape is
quite robust and agrees well with the behavior found in the data.  At
leading order in $v$, results for $\Upsilon(2S)$ can be obtained
simply by changing $|R_{\Upsilon(nS)}(0)|^2$ and the branching
ratio. The predictions for the polarization parameter $\alpha$ are
only slightly affected by the NNLO$^\star$ contributions and remain
negative at high transverse
momentum,~\cf{polarizationalphas4}. Finally, we show our predictions
for the $p_T$ distributions at the LHC ($\sqrt{s}=14$ TeV)
in~\cf{upsilon1S14}.

Obviously, our analysis cannot be extended to rather low $p_T$, where the
approximations on which it is based no longer hold. In this case, one
could improve the predictions by merging the matrix elements with
parton showers using one of the approaches available in the
literature~\cite{Alwall:2007fs}, or by performing an
analytic resummation~\cite{Berger:2004cc}.

\begin{figure}[ht!]
\centering
\includegraphics[width=0.95\columnwidth,angle=0]{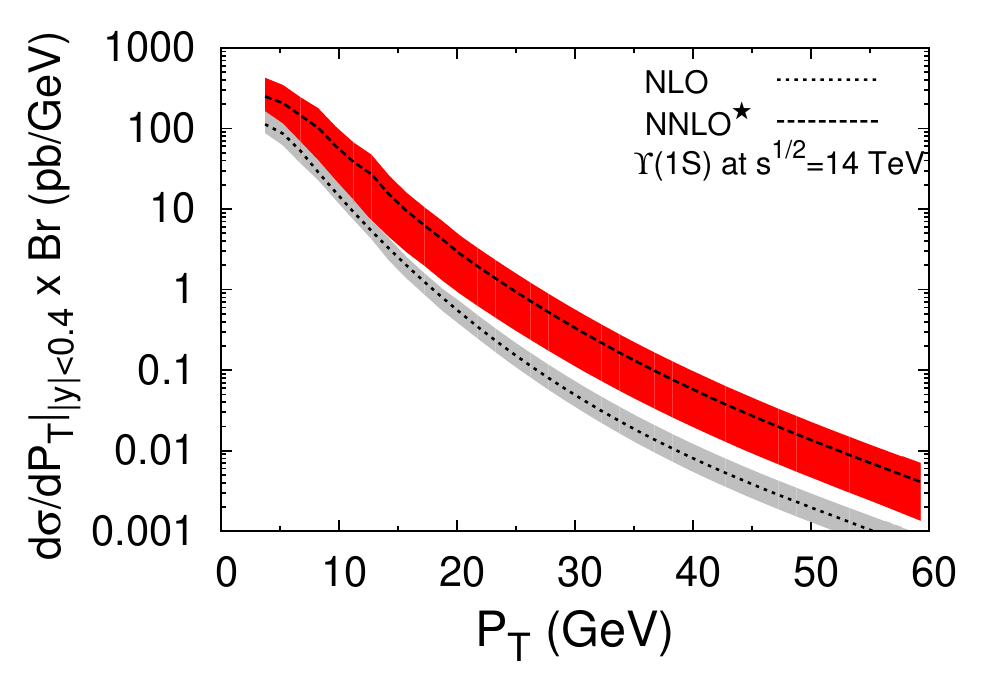}
\caption{$\Upsilon(1S)+X$ at $\sqrt{s}=14$ TeV. Same parameters as for Fig 2. 
Results for the $\Upsilon(3S)+X$ can be obtained by rescaling the curves 
by the ratio of the corresponding wave functions values at the origin and
branching ratios.}
\label{upsilon1S14}
\end{figure}


In conclusion, we have compared the full NLO ($\alpha_S^4$) cross
section for polarized direct hadroproduction of $\Upsilon(nS)$ to the
data from the Tevatron. At this order, the polarization parameter $\alpha$
is negative and decreases with $p_T$. However, the yield is still too
low compared to the data especially at high $p_T$.

This led us to consider a subset of the NNLO ($\alpha_S^5$)
corrections, namely those corresponding to the production of a
$\Upsilon$ associated with three light partons and containing
$p_T^{-4}$ contributions. Integrating the amplitudes for such
processes at $\alpha_S^5$ leads to divergences in some phase-space
regions that we avoided by imposing a minimal invariant mass between
the light partons.  This was motivated by the fact that the neglected
contributions are proportional to $\alpha_S^4$ amplitudes, which scale
at most as $p_T^{-6}$, so that we expected this approach to give a
reliable estimate of the full $\alpha_S^5$ result at large $p_T$. We
validated our expectations by comparing such an approach with the
computation at $\alpha_S^4$ and by explicitly checking the sensitivity
of our results to the cutoff choice at high $p_T$. The estimated
$\alpha_S^5$-contributions appear to fill the gap between the full NLO
($\alpha_S^4$) and the data.

As a further result, we have predicted the polarization parameter
$\alpha$ to be negative and only slightly decrease with $p_T$,
indicating a longitudinally polarized yield. More accurate data, in
particular for $\Upsilon(3S)$ and possibly at higher $p_T$, are needed 
to check these predictions.

In summary, the comparison with the available data suggests that there is 
no need to include contributions from color-octet transitions for 
$\Upsilon(nS)$ production. For the $\psi$ case, the situation 
is less clear since the addition of higher-order QCD corrections seems not
to suffice to describe the data~\cite{Campbell:2007ws,Gong:2008sn,Gong:2008ft}. In this case, other mechanisms may be at work such as those recently discussed in Refs.~\cite{Haberzettl:2007kj,Nayak:2007mb}.


\vspace*{-.6cm}

\begin{thebibliography}{99}

\bibitem{review}
  J.\,P.~Lansberg, Int.\ J.\ Mod.\ Phys.\ A {\bf 21} (2006) 3857;
N.~Brambilla {\it et al.}, CERN 2005-005, hep-ph/0412158;
  M.~Kramer,
  Prog.\ Part.\ Nucl.\ Phys.\  {\bf 47} (2001) 141.

\bibitem{CSM_hadron}
C-H. Chang,
{Nucl. Phys. } B {\bf 172} (1980) 425;
R. Baier and R. R\"uckl,
{Phys. Lett. } B {\bf 102} (1981) 364;
{Z. Phys. } C {\bf 19} (1983) 251;
E.~L.~Berger and D.~L.~Jones,
Phys.\ Lett.\  B {\bf 121} (1983) 61.





\bibitem{Bodwin:1994jh}
  G.~T.~Bodwin, E.~Braaten and G.~P.~Lepage,
  Phys.\ Rev.\ D {\bf 51}, 1125 (1995)
  [Erratum-ibid.\ D {\bf 55}, 5853 (1997)]


\bibitem{Affolder:1999wm}
  A.~A.~Affolder {\it et al.}  [CDF Collaboration],
  Phys.\ Rev.\ Lett.\  {\bf 84} (2000) 2094

\bibitem{Acosta:2001gv}
  D.~Acosta {\it et al.}  [CDF Collaboration],
  Phys.\ Rev.\ Lett.\  {\bf 88} (2002) 161802.


\bibitem{Braaten:2000cm}
  E.~Braaten, S.~Fleming and A.~K.~Leibovich,
  Phys.\ Rev.\  D {\bf 63} (2001) 094006.

\bibitem{D0note}
  V.~M.~Abazov {\it et al.}  [D0 Collaboration],
  $D\emptyset$ Note 5089-conf


\bibitem{Campbell:2007ws}
  J.~Campbell, F.~Maltoni and F.~Tramontano,
  Phys.\ Rev.\ Lett.\  {\bf 98} (2007) 252002.


\bibitem{Artoisenet:2007xi}
  P.~Artoisenet, J.~P.\,Lansberg and F.~Maltoni,
  Phys.\ Lett.\  B {\bf 653} (2007) 60.


\bibitem{Pumplin:2002vw}
  J.~Pumplin, D.~R.~Stump, J.~Huston, H.~L.~Lai, P.~Nadolsky and W.~K.~Tung,
  JHEP {\bf 0207} (2002) 012


\bibitem{Abazov:2005yc}
  V.~M.~Abazov {\it et al.}  [D0 Collaboration],
  Phys.\ Rev.\ Lett.\  {\bf 94} (2005) 232001


\bibitem{Bodwin:2007fz}
  G.~T.~Bodwin, H.~S.~Chung, D.~Kang, J.~Lee and C.~Yu,
  Phys.\ Rev.\  D {\bf 77} (2008) 094017.



\bibitem{kt}
  Ph.~Hagler~{\it et al.} 
  Phys.\ Rev.\  D {\bf 63} (2001) 077501;
  F.~Yuan and K.~T.~Chao,
  Phys.\ Rev.\  D {\bf 63} (2001) 034006.

\bibitem{Artoisenet:2007qm}
  P.~Artoisenet, F.~Maltoni and T.~Stelzer,
  JHEP {\bf 0802} (2008) 102.

\bibitem{Alwall:2007fs}
  J.~Alwall {\it et al.},
  Eur.\ Phys.\ J.\  C {\bf 53} (2008) 473.

\bibitem{Berger:2004cc}
  E.~L.~Berger, J.~w.~Qiu and Y.~l.~Wang,
  Phys.\ Rev.\  D {\bf 71} (2005) 034007

\bibitem{Gong:2008sn}
  B.~Gong and J.~X.~Wang,
  arXiv:0802.3727 [hep-ph].

\bibitem{Gong:2008ft}
  B.~Gong, X.~Q.~Li and J.~X.~Wang,
  arXiv:0805.4751 [hep-ph].

\bibitem{Haberzettl:2007kj}
H.~Haberzettl and J.~P.~Lansberg,
Phys.\ Rev.\ Lett.\ {\bf 100} (2008) 032006;
  J.~P.~Lansberg, J.~R.~Cudell and Yu.~L.~Kalinovsky,
  Phys.\ Lett.\  B {\bf 633} (2006) 301


\bibitem{Nayak:2007mb}
  G.~C.~Nayak, J.~W.~Qiu and G.~Sterman,
  Phys.\ Rev.\ Lett.\  {\bf 99} (2007) 212001



\end{thebibliography}
\end{document}